\documentclass[3p,times,final]{elsarticle}
\usepackage{ecrc}
\usepackage[bookmarks=false]{hyperref}
    \hypersetup{colorlinks,
      linkcolor=blue,
      citecolor=blue,
      urlcolor=blue}
\usepackage{amsmath}
\usepackage{xparse}
\usepackage{wrapfig}
\usepackage{enumitem}
\volume{00}
\usepackage[skip=4pt]{caption}
\firstpage{1}
\journal{Procedia Computer Science}
\usepackage{titlesec}
\titlespacing*{\section}{0pt}{8pt plus 2pt minus 2pt}{4pt}
\titlespacing*{\subsection}{0pt}{6pt plus 2pt minus 2pt}{3pt}
\runauth{G.~Aarts et al.}

\jid{procs}

\usepackage{amssymb}

\makeatletter
\def\@oddfoot{}
\def\@evenfoot{}
\makeatother

\usepackage{xcolor}
\usepackage{subcaption} 

\newif\ifshowcomments
\showcommentsfalse

\newcommand{\ar}[1]{\ifshowcomments \textcolor{red}{{\bfseries AR: }#1 }\\ \fi}

\begin{document}
\begin{frontmatter}

\dochead{Proceedings of the fourth EuroHPC User Days 2026}%%%
%% Use \dochead if there is an article header, e.g. \dochead{Short communication}
%% \dochead can also be used to include a conference title, if directed by the editors
%% e.g. \dochead{17th International Conference on Dynamical Processes in Excited States of Solids}
\title{Future Requirements of Lattice Field Theory Calculations on European High-Performance Computing Facilities}
%\title{Future requirements of lattice field theory calculations on European High Performance Computing Facilities}

\author[a]{Gert Aarts}
\author[b]{Gunnar Bali}
\author[c]{Jacob Finkenrath}
\author[d,e]{Stefan Krieg}
%\author[d]{Stefan Krieg}
\author[f]{Antonio Rago\corref{cor1}}
\ead{rago@qtc.sdu.dk}
\author[e]{\mbox{Carsten Urbach}}
\author[g]{Hartmut Wittig}

\address[a]{Centre for Quantum Fields and Gravity, Department of Physics, Swansea University, Swansea, SA2 8PP, United Kingdom}
\address[b]{Institut für Theoretische Physik, Universität Regensburg, 93040 Regensburg, Germany}
\address[c]{Department of Physics, Bergische Universität Wuppertal, 42119 Wuppertal, Germany}
\address[d]{Jülich Supercomputing Centre (JSC), Center for Advanced Simulation
and Analytics (CASA), Forschungszentrum Jülich, 54245 Jülich, Germany}
\address[e]{Helmholtz-Institut für Strahlen- und Kernphysik, Rheinische Friedrich-Wilhelms-Universität Bonn, Nussallee 14-16, 53115 Bonn, Germany}
\address[f]{Quantum Theory Center (QTC), IMADA, University of Southern Denmark, Campusvej 55, 5230 Odense M, Denmark}
\address[g]{PRISMA++ Cluster of Excellence, Institut für Kernphysik and Helmholtz-Institut Mainz, Johannes Gutenberg-Universität
Mainz,\newline 55099 Mainz, Germany}
\begin{abstract}
Lattice field theory provides a first-principles framework for studying properties of strongly interacting quantum field theories in elementary particle physics.
Researchers in lattice field theory are also among the largest and most efficient users of high-performance computing resources in fundamental science. In this contribution, we outline the computational profile of lattice QCD, from gauge-field generation to large-scale measurements, and discuss the main hardware, software, and human resource requirements needed to sustain progress on current and future European HPC infrastructures.
\end{abstract}


%\correspondingauthor[*]{Corresponding author. Tel.: +0-000-000-0000 ; fax: +0-000-000-0000.}
\cortext[cor1]{Corresponding author. Tel.: +4565509258}
\end{frontmatter}

%%
%% Start line numbering here if you want
%%
% \linenumbers
\vspace*{6pt}
%% main text

\section{Introduction}
\label{sec:intro}

%\subsection{Physics motivation and achievements}

The Standard Model of particle physics represents one of the most successful theoretical frameworks in modern science.
% High Energy Physics
Major international experiments carried out over several decades have established with remarkable precision the particle content and interactions of the electroweak and strong sectors. Measurements at facilities such as LEP, the Tevatron, and the Large Hadron Collider have confirmed the gauge structure of the Standard Model and determined many of its parameters with high accuracy.

Within this framework, the strong interaction described by Quantum Chromodynamics (QCD) plays a central role. QCD governs the dynamics of quarks and gluons and is responsible for the formation of hadrons, including baryons, the particles that make up most of the visible matter in the universe. A key feature of this theory is that the interaction becomes stronger at low energies. At the scale relevant for hadrons, the theory is therefore strongly coupled, meaning that standard perturbative techniques are no longer applicable. Understanding the properties of strongly interacting matter consequently requires genuinely non-perturbative approaches.

Lattice field theory provides such a framework. In this approach space and time are discretised on a finite grid, allowing the quantum field theory to be formulated in a way that can be evaluated numerically. Physical observables are computed by sampling gauge field configurations using Monte Carlo methods. This formulation enables calculations directly from the fundamental theory with systematically improvable uncertainties.
Although originally developed for QCD, the method can be applied to any quantum field theory that admits a Euclidean path integral formulation and therefore has become a powerful tool for studying any strongly interacting field theory.

Lattice Quantum Chromodynamics represents the most mature realisation of this approach. Over the past decades lattice QCD has produced a number of major achievements. One of the earliest successes was the quantitative demonstration of colour confinement through the computation of the potential between a heavy quark and an antiquark, showing the emergence of a constant attractive force at large distances.
%linearly rising interaction at large distances.
Lattice calculations have also enabled precise determinations of the hadron spectrum, reproducing the masses of light hadrons with percent-level agreement with experimental measurements.

In recent years, lattice QCD has become an essential tool for determining fundamental parameters of the Standard Model, including the quark masses and the strong coupling constant. Its scope has expanded to increasingly complex observables, such as nucleon structure, hadron scattering amplitudes, and
electroweak decay matrix elements.
% removed the glueballs because this is no expansion. The first
% calculations were glueballs.
Lattice calculations also provide crucial theoretical input for precision tests of the Standard Model. A notable example is the determination of the hadronic contributions to the anomalous magnetic moment of the muon $(g-2)$, a quantity that showed a tension between experimental measurements and theoretical predictions \cite{Aliberti:2025beg}. 
%Improving the precision of hadronic contributions has therefore been essential for interpreting this discrepancy and assessing possible indications of physics beyond the Standard Model 
%
Improving the precision of hadronic contributions is therefore essential for interpreting this discrepancy and assessing possible indications of physics beyond the Standard Model (BSM) \cite{deBlas:2025xxx}. 
Including temperature in simulations is straightforward and yields results relevant for the early universe and for relativistic heavy-ion collisions, e.g., at the Large Hadron Collider \cite{Aarts:2023vsf}.
The community has embraced this challenge in line with the priorities identified by the European Strategy for Particle Physics\footnote{\url{https://europeanstrategy.cern/}} and the Nuclear Physics Collaboration Committee (NuPECC)
Long Range Plan\footnote{\url{https://www.nupecc.org/}}.
%
%This challenge is 
%The challenges encountered are not limited to QCD only. Many extensions of the Standard Model predict the existence of additional strongly interacting sectors, often motivated by attempts to explain phenomena such as electroweak symmetry breaking, dark matter, or the hierarchy problem. These Beyond the Standard Model (BSM) theories frequently involve new gauge dynamics whose low-energy behaviour is inherently non-perturbative. As a result, the ability to study strongly coupled quantum field theories from first principles is essential both for interpreting present experimental data and for exploring possible new physics scenarios.
%
%These developments have established lattice field theory as a mature computational discipline providing quantitative predictions for strongly interacting systems directly from the underlying quantum field theory, while at the same time offering a versatile framework for exploring strongly coupled dynamics that may arise in theories beyond the Standard Model.

The challenges encountered are not limited to QCD. Many extensions of the Standard Model predict additional strongly interacting sectors, often motivated by attempts to explain phenomena such as electroweak symmetry breaking, dark matter, or the hierarchy problem. These BSM theories frequently involve new gauge dynamics whose low-energy behaviour is inherently non-perturbative. In summary, lattice field theory has become a mature computational discipline for obtaining quantitative predictions for strongly interacting systems directly from the underlying quantum field theory, while also providing a versatile first-principles framework for interpreting present experimental data and exploring possible new physics scenarios beyond the Standard Model.

\subsection{Lattice formulation}
The lattice formulation replaces continuous space--time with a discrete grid characterised by a lattice spacing $a$ and a finite physical extent $L$. The lattice spacing acts as an ultraviolet regulator of the theory, while the finite volume provides an infrared cutoff.

Physical observables are computed as expectation values of operators defined on the lattice. In practice, these expectation values are estimated by averaging over ensembles of gauge-field configurations. Each configuration can be interpreted as a snapshot of the quantum fields over the entire space--time lattice. Since quantum fluctuations are intrinsic to the theory, a single configuration does not carry direct physical meaning. Instead, reliable predictions emerge only after sampling a sufficiently large number of configurations distributed according to the probability measure defined by the lattice action.

%These configurations are generated using
Markov-chain Monte Carlo algorithms produce a sequence of such field configurations.
%sampling the path integral distribution. (path integrals may be confusing for the non-expert)
Observables are then evaluated on each configuration (``measurements'') and averaged over the ensemble. The statistical uncertainty associated with this procedure decreases as more configurations are included, allowing one to assign predictive errors to the final results. Controlling these statistical uncertainties requires careful treatment of correlations between configurations and different observables on a given ensemble.
%often demands very large ensembles.

In addition to statistical uncertainties, lattice calculations must address systematic effects introduced by the discretisation of space--time. The physical theory corresponds to the continuum limit, which is obtained by extrapolating results computed at several lattice spacings towards $a \to 0$, a procedure called ``continuum limit''. Achieving control over this extrapolation is challenging and requires simulations at multiple lattice spacings together with careful analysis of discretisation effects. At the same time, the lattice volume must be large enough to suppress finite-volume artefacts, and the parameters of the theory must be tuned to reproduce some observables of reference.

Reducing the lattice spacing, increasing the lattice volume, and approaching physical quark masses all lead to rapidly increasing computational costs. As a result, lattice QCD simulations require extremely large computational resources and have historically been among the most demanding applications in high-performance computing.

\subsection{Lattice field theory community}
\ar{shall we mention the size of the community?}
A key feature of lattice field theory is that physical results, once extrapolated to the continuum limit, do not depend on the specific lattice formulation used. Different discretisations of the theory may differ at finite lattice spacing, but they all converge to the same continuum physics. This property allows for independent calculations of the same observables using different lattice formulations. In practice, this provides an important and, in fact, scientifically necessary form of cross-validation of the method: results obtained by different groups using different discretisations can be compared and tested against each other. In this respect, lattice QCD resembles large experimental programmes such as those at CERN, where multiple experiments independently measure the same quantities using different detectors to ensure robustness and reliability of the results.

The choice of a particular lattice formulation often shapes the organisation of the research collaborations that carry out the simulations. Over time, communities have formed around specific discretisations of the theory, together with the software frameworks and algorithmic strategies required to implement them efficiently. These collaborations typically involve multiple research institutions and national computing infrastructures, pooling expertise in physics, numerical algorithms, and high-performance computing.

The lattice field theory community, therefore, operates through large collaborative projects that generate and analyse ensembles of gauge-field configurations. Producing such ensembles requires extensive and sustained access to high-performance computing resources and is frequently coordinated across several computing centres. Once generated, these configurations become long-lived scientific assets within an open-science framework, shared across collaborations, and reused by multiple groups for diverse physics analyses, retaining and even increasing their value over more than a decade.
%
%Because the same ensembles support many independent studies, the community has historically developed a strongly interconnected structure. This model maximises the scientific return of large computational investments and has fostered a culture of data sharing, common standards, and joint development of software tools that underpin modern lattice field theory research.

Because the same ensembles support many independent studies, the community has historically developed a strongly interconnected structure. This model maximises the scientific return of large computational investments and has fostered a culture of data sharing, common standards, and joint development of software tools that underpin modern lattice field theory research. It also enables the community to scrutinise results obtained with different lattice formulations and to provide validated input to the wider community, in particular flavour physics, where lattice QCD supplies the non-perturbative hadronic quantities needed to interpret precision measurements from LHCb, Belle II, BESIII, kaon experiments, and related programmes. Decay constants, form factors, bag parameters, and mixing matrix elements enter directly into CKM phenomenology, unitarity tests, and searches for physics beyond the Standard Model. The Flavour Lattice Averaging Group (FLAG) provides a central example of this community-wide validation process: it reviews lattice results, applies agreed quality criteria, and provides reference averages or estimates for the wider particle-physics community \cite{FlavourLatticeAveragingGroup:2019iem,FlavourLatticeAveragingGroupFLAG:2021npn,FlavourLatticeAveragingGroupFLAG:2024oxs}.
\subsection{EuroLFT}
The European lattice field theory community has recently established the EuroLFT initiative (\url{https://eurolft.github.io/}) as a platform to represent the interests and needs of the European lattice community towards European institutions. EuroLFT acts as a point of reference for institutions seeking to engage with the lattice field theory community and facilitates dialogue on topics relevant to research, computing infrastructure, and long-term scientific strategy. Similar initiatives exist in other regions of the world; for example, in the United States the lattice community is organised through the USQCD collaboration, which coordinates community activities and interactions with national laboratories and funding agencies (\url{https://www.usqcd.org}).
%
%EuroLFT aims to strengthen connections within the European lattice community and to improve interactions between research groups, computing centres, and policy bodies. Its role is to articulate the computational requirements of the field, provide a collective perspective on future developments in computing architectures, and contribute community input to discussions shaping European research infrastructures.
%
%By serving as a representative forum for the community, EuroLFT aims to ensure that the specific needs of lattice field theory are visible in broader strategic discussions concerning high-performance computing and scientific data infrastructures in Europe.

EuroLFT aims to strengthen connections within the European lattice community and to improve interactions between research groups, computing centres, and policy bodies. By serving as a representative forum for the community, it articulates the computational requirements of the field, provides a collective perspective on future developments in computing architectures, and ensures that the specific needs of lattice field theory are visible in broader strategic discussions concerning high-performance computing and scientific data infrastructures in Europe.

\subsection{International Lattice Data Grid}
The lattice field theory community has historically embraced the principles of open science. A central element of this approach is the sharing of gauge-field configurations, which represent a major fraction of the consumed computer time 
%the most computationally expensive output
of large-scale lattice simulations. Once generated, these configurations can be reused by many research groups to study a wide range of physical observables. This practice greatly amplifies the scientific return of large computational investments and is consistent with the culture of collaboration and data sharing across the field.

To support this model, the community established the International Lattice Data Grid (ILDG)~\cite{Davies:2002mu,Irving:2003uk,Ukawa:2004he,Beckett:2009cb}, a distributed service to share gauge-field ensembles produced by lattice collaborations worldwide. The ILDG connects multiple regional storage nodes through a common metadata catalogue, enabling researchers to discover, access, and reuse configurations generated on major high-performance computing systems.
The ILDG represented, and still represents, an implementation of FAIR principles~\cite{wilkinson2016fair} for the precious lattice gauge-field ensembles.

More recently, the European part of the community has launched the ILDG\,2.0 initiative~\cite{Karsch:2022tqw}, aimed at revitalising this infrastructure by adapting it to modern data management practices and evolving computing environments. This effort seeks to improve interoperability with contemporary storage technologies and data services while preserving the core mission of enabling open access to lattice simulation data. In addition, ILDG\,2.0 strives to extend its service to additional data types.

\section{Computational profile}
The computational workflow of lattice QCD spans several stages, each characterised by different computational patterns. At its core lies the generation of gauge-field configurations using large-scale Monte Carlo simulations. These configurations form ensembles that are subsequently used for a wide range of physics studies.

Once generated, ensembles often remain scientifically valuable for many years. New analysis techniques are continuously applied to existing configurations and novel observables measured, meaning that the computational investment in ensemble generation continues to produce results over long timescales.

%Different discretisations of the QCD action may be used to approach the continuum limit. Although these formulations differ at finite lattice spacing, they converge to the same physical predictions in the continuum limit due to universality. This allows different collaborations to pursue complementary computational strategies while ultimately addressing the same physical theory.
% Have commened out because this is repeated!

\subsection{Typical workflows}
Lattice workflows can involve a wide variety of computational tasks and analysis steps, often combining several software frameworks and numerical techniques. In practice, the full scientific pipeline may include configuration generation, ``measurements'', and increasingly sophisticated analysis procedures. In this section we provide a simplified overview of these workflows, focusing on the main components that dominate the overall computational cost.

{\bfseries Configuration generation.} 
Gauge configurations are generated using the Hybrid Monte Carlo (HMC) algorithm \cite{Duane:1987de} and related methods that sample the probability distribution defined by the lattice action as a Markov chain. In practice, the simulation evolves a gauge field configuration through a sequence of molecular-dynamics trajectories, with a Metropolis accept/reject step after each individual trajectory.
The size of the lattice is determined by two main physical requirements. The lattice spacing $a$ must be small enough to control discretisation effects, while the physical volume $\propto L^4$ must be large enough to accommodate the relevant hadronic states without significant finite-volume distortions. 
These constraints are closely related to the characteristic physical scales that must be resolved in the simulation. At long distances, the box size must be large enough to accommodate the lightest hadronic state in the theory, the pion. At the same time, the lattice spacing must be small enough to resolve the shortest physical scales relevant for the observables under investigation, which are typically set by the heaviest states or largest momenta entering the problem. 

For present-day calculations with close to physical quark masses, this typically leads to lattice spacings in the range $a \sim 0.04$--$0.1$ fm and spatial volumes with linear sizes of several fermis ($1\ \mathrm{fm} = 10^{-15}\ \mathrm{m}$). As a result, typical lattice grids used in modern simulations contain between $10^7$ and $10^{9}$ lattice sites, corresponding, for example, to lattices of size $64^3 \times 128$, $96^3 \times 192$, or $128^3\times 256$.

The computational cost of this procedure is dominated by repeatedly solving  large sparse linear systems involving the lattice Dirac operator, which describes the propagation of quarks in the background gauge field. These linear systems typically involve matrices with dimensions proportional to the number of lattice sites multiplied by internal spin and colour degrees of freedom.
For the lattice volumes quoted above this amounts to order $10^{10}$ degrees of freedom in the linear system, or more.

%In addition, as the lattice spacing becomes smaller, the evolution of the Markov chain becomes increasingly slow. In particular the global topological charge of the gauge field may evolve only very rarely, a phenomenon known as topological freezing. This effect leads to long autocorrelation times and requires molecular-dynamics trajectories to be extended over increasingly long simulation chains in order to obtain statistically independent configurations \cite{Schaefer:2010hu}.
% different strategies exist. I shortened this.

In addition, as the lattice spacing is decreased, autocorrelation times grow rapidly~\cite{DelDebbio:2004xh,Schaefer:2010hu} and increasingly long Markov chains need to be generated in order to obtain a sufficiently large number of statistically independent configurations. This effect, combined with large lattice volumes required to control finite-volume effects, makes configuration generation one of the most computationally demanding components of lattice QCD simulations and of workloads in high-performance computing in general. \ar{can we quote some number?}

While the discussion above provides a useful estimate of the computational cost of lattice QCD simulations, it represents only a simplified picture. Over the past decades the lattice community has continuously worked to overcome these limitations through the development of improved numerical algorithms alongside advances in computing hardware.

Key developments include refinements of the Hybrid Monte Carlo algorithm \cite{Duane:1987de}, such as mass-precon\-di\-tioning \cite{Hasenbusch:2001ne}, domain-decomposition techniques \cite{Luscher:2003qa}, deflation methods \cite{Luscher:2007se}, and, more recently, multigrid solvers for the lattice Dirac operator \cite{Babich:2010qb}. These innovations have significantly reduced the cost of the most demanding parts of the calculation and have enabled simulations at or near the physical quark masses, at small lattice spacings.

{\bfseries Measurements of observables.} 
Once gauge-field configurations have been generated, physical observables are extracted through ``measurement calculations'' performed on these ensembles. Most observables in lattice QCD are obtained from correlation functions constructed from quark and gluon fields. In practice, computing these correlation functions requires solving additional linear systems involving the Dirac operator, similar to those encountered during configuration generation.
These measurements typically require many independent inversions of the Dirac operator for each configuration. To improve statistical precision, the calculations are repeated for multiple source locations on the lattice and often for several operator structures.
%Averaging over many measurements reduces statistical fluctuations and allows the extraction of physical quantities such as hadron masses, matrix elements, or form factors.
% some repetition here

Observables of interest such as hadron masses, matrix elements, or form factors are extracted from correlation functions at large Euclidean times.
A central challenge in these calculations is the signal-to-noise problem. For many observables, particularly those involving baryons or multi-hadron systems, the signal contained in correlation functions decreases exponentially with the Euclidean time separation, while the statistical noise decreases much more slowly. As a result, the ratio between signal and noise deteriorates exponentially at large time separations, making it increasingly difficult to extract the desired physical information.

Controlling this effect requires ensembles of large numbers of configurations and a significant number of measurements per configuration. Various strategies are used to mitigate the problem, including improved operator constructions, variance-reduction techniques, and stochastic estimation methods. Nevertheless, signal-to-noise degradation remains a fundamental limitation that often drives the overall computational cost of many lattice QCD studies.

{\bfseries Machine-learning assisted workflows.}  
Recent developments have introduced machine-learning techniques into several aspects of lattice QCD workflows. These approaches are being explored for tasks along the entire pipeline of a lattice QCD simulation: from the tuning of the action parameters, via the generation of ensembles and the computation of observables, to the data analysis \cite{Boyda:2022nmh}. 
As an example, it is now possible to machine-learn a renormalisation--group-improved lattice gauge action (``perfect action'') \cite{Holland:2025fsa} via the use of {\em lattice gauge equivariant convolutional neural networks} \cite{Favoni:2020reg}.
Variance or noise reduction in correlation functions can be achieved via the use of (automated) differentiation of lower $n$-point functions \cite{Catumba:2025ljd,Abbott:2026ylv}.

A particularly active area of research concerns the use of generative models to improve sampling in Monte Carlo simulations. In this context, methods in generative AI are being adopted as a way to learn transformations that map simple probability distributions to the highly structured distributions describing lattice field configurations. 
A distinction has to be drawn between approaches that learn from the action, such as normalising flow \cite{Cranmer:2023xbe}, and those that learn from data, such as diffusion models \cite{Aarts:2026zzr}, and hence rely on large-scale ensembles generated using standard (HMC) approaches. 
Once trained, such models can potentially generate candidate configurations with reduced autocorrelations, or provide improved proposal distributions for Monte Carlo algorithms. These approaches may help mitigate some of the algorithmic challenges associated with critical slowing down,
%and topological freezing (deleted because also removed above)
which increasingly affect simulations at fine lattice spacings. While there are some attempts to go to four dimensions \cite{Abbott:2025kvi}, so far the emphasis is on lower-dimensional models. Importantly, there are deep links between methods proposed in the machine-learning literature and those developed in lattice field theory, which provides a fruitful path for future interaction \cite{Caselle:2022acb,Wang:2023exq}.
%It is expected that these methods will develop quickly and that large computing resources will become essential.
A strategic roadmap, based on a survey across particle and nuclear physics, is now available \cite{Caron:2025rir}.

Machine-learning techniques are also being explored as surrogate models capable of approximating expensive components of lattice calculations, thereby reducing the number of costly solver applications required in certain analysis tasks. While these approaches remain at an exploratory stage and do not yet replace the established numerical algorithms used in lattice simulations, they offer promising avenues for improving efficiency and scalability. 
As modern HPC architectures increasingly integrate AI-oriented hardware and software ecosystems, machine-learning-assisted methods may become an important complement to traditional lattice QCD workflows.  

\subsection{Main hardware bottlenecks}
Modern lattice QCD codes are dominated by repeated applications of sparse matrices, such as the Dirac operator, within iterative linear solvers. While the computational effort is perfectly balanced over the space-time grid, the corresponding kernels have relatively low arithmetic intensity and are, therefore, often limited by memory bandwidth and communication rather than peak floating-point performance. Modern multigrid and deflated solvers further amplify this effect, as key components of these algorithms reduce the local arithmetic workload and make performance increasingly dependent on low-latency, high-bandwidth network communication.
%
%This is even more the case as modern multi-grid solvers have reduced the arithmetic intensity further.
%
%Thus, high-bandwidth and low-latency interconnects between compute units are key requirements for lattice QCD and lattice field theory applications targeted at breakthrough calculations that answer open questions in nuclear, hadron, and particle physics. In particular, for measurement applications, large CPU or GPU local memory is of utmost importance, as it helps to significantly increase the multiple usage of single solutions to the Dirac equation.
Thus, high-bandwidth, low-latency interconnects and large local CPU/GPU memory are key requirements, especially for measurement applications where solutions of the Dirac equation are reused extensively.

At the same time, increasing floating-point precision is becoming an important requirement. Although mixed-precision algorithms are often used in inner iterations of solvers to maximise performance, many observables require high numerical accuracy and strict control of rounding errors. This requirement becomes more pronounced as lattice volumes grow, where maintaining numerical stability over large configuration volumes becomes essential. In this context, current trends in GPU hardware, which increasingly prioritise lower-precision arithmetic to maximise throughput, could be suboptimal for lattice workloads, where sustained and reliable double-precision performance remains critical. A potential solution may be the use of emulated high-precision arithmetic. This would, however, require a reliable and verified implementation of the emulation for simulation results to be trustworthy.  

\subsection{Evolution of computational effort}
%\begin{figure}[t!]
%    \centering
%    \includegraphics[width=0.9\textwidth]{data_euroHPC.pdf}
%    \caption{Fraction of allocated resources in regular (left) and extreme (right) EuroHPC calls. Computational physics allocations 
    %are entirely accounted for by lattice field theory.}
%    \label{fig:Eurohpc_lattice}
%\end{figure}
\begin{wrapfigure}{r}{0.4\linewidth}
    \centering
    \includegraphics[width=\linewidth]{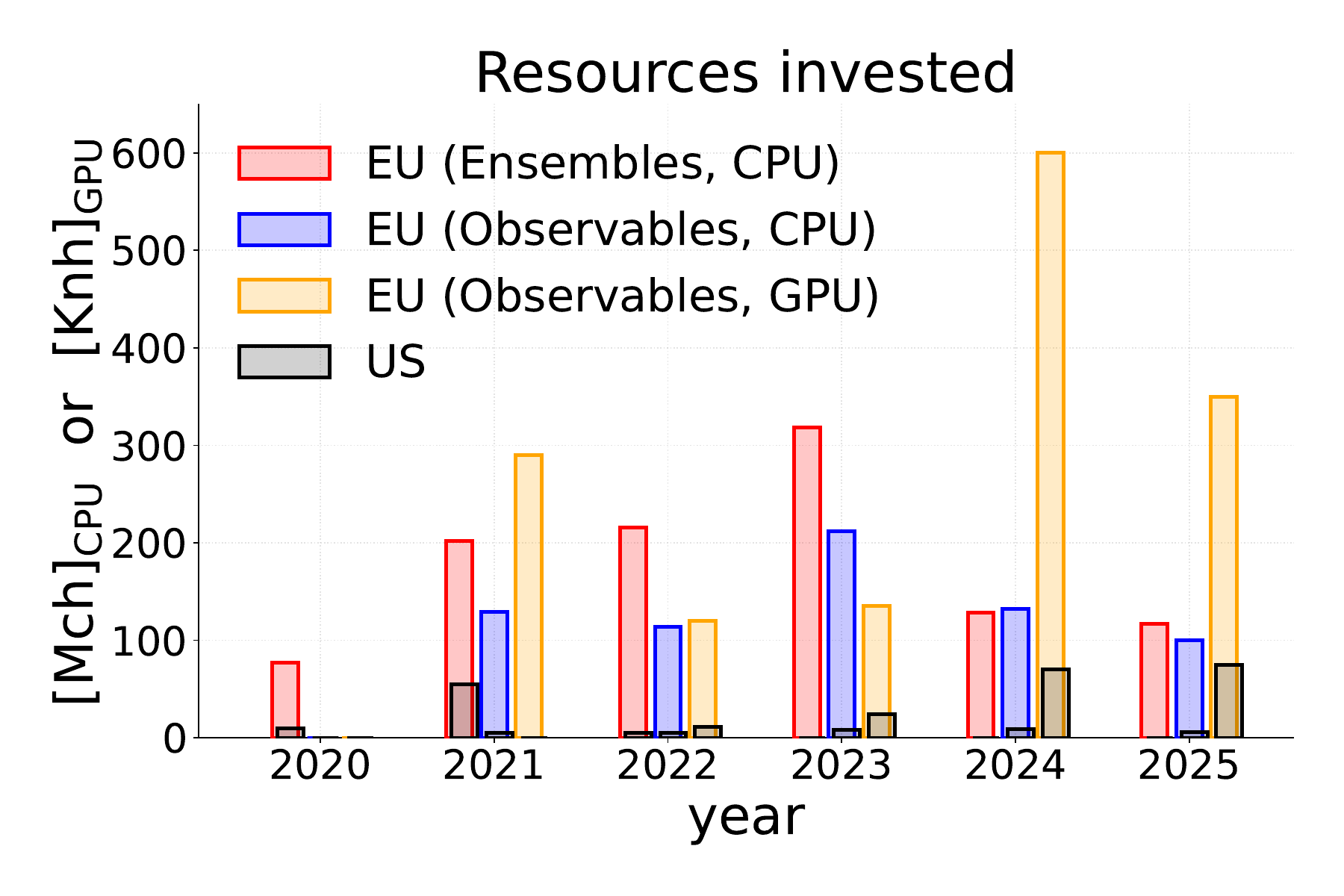}
    \caption{Evolution of computational resources invested in the OpenLAT collaboration, separated into configuration generation and observable measurements.}
    \label{fig:openlat_cost}
\end{wrapfigure}
Over the past decades the computational scale of lattice QCD simulations has grown dramatically. %Early simulations were performed on small lattice volumes and modest computing resources. 
Early simulations were limited to comparatively small lattice volumes, using computing resources that were state of the art at the time but modest by present-day standards. Modern simulations routinely employ volumes exceeding $64^3 \times 192$ with near-physical quark masses \cite{Szabo:2024pjz,Alexandrou:2025bkm} and require large statistics \cite{Borsanyi:2025ygf}.
%Large collaborations that are based in Europe or have a European component such as CLS, CSSM/QCDSF/UKQCD, ETMC, FASTSUM, HotQCD, openLAT, RBC/UKQCD, RC*, TELOS and TWEXT coordinate ensemble production campaigns across multiple HPC systems. Individual ensembles may require hundreds of millions of CPU core-hours or hundreds of thousands of GPU hours to generate. Each ensemble subsequently supports numerous physics analyses, maximising the scientific return on computational investments.
A non-exhaustive list of large collaborations based in Europe, or with a significant European component, includes CLS, CSSM/QCDSF/UKQCD, ETMC, FASTSUM, HotQCD, openLAT, RBC/UKQCD, RC*, TELOS, and TWEXT. These collaborations coordinate ensemble production campaigns across multiple HPC systems and the subsequent sharing of the resulting ensembles.

\begin{figure}[t!]
    \centering
    \includegraphics[width=0.9\linewidth]{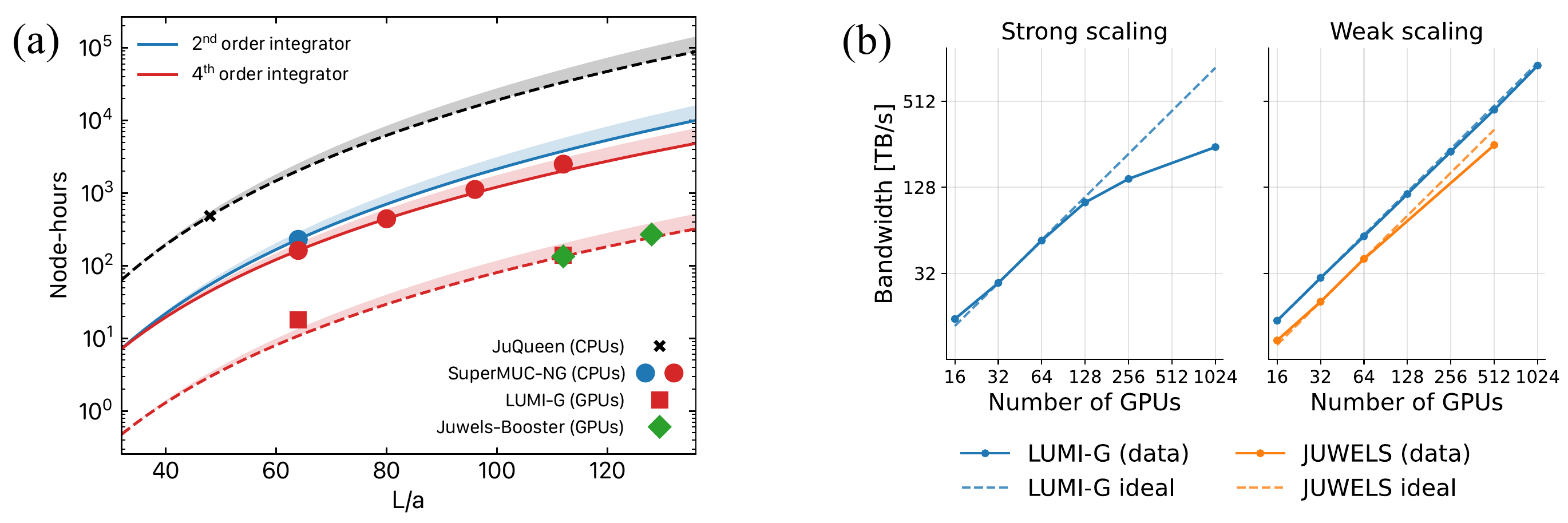}
    \caption{(a): \texttt{tmLQCD}: Cost per molecular dynamics unit for a simulation with light, strange and charm quarks tuned to physical masses. Utilising multi-grid solvers accelerated the simulations by a factor~10. Using GPUs decreases the required node-hours by another factor~10. (b): \texttt{HiRep}: Strong and weak scaling of the Dirac operator on LUMI-G and JUWELS Booster with all dimensions parallelised up to 1024 GPUs. %for an SU(3) gauge group with fundamental fermions.
    }
    \label{fig:cost_alg_gpu}
\end{figure}

As an illustrative example, fig.~\ref{fig:openlat_cost} shows the evolution of computational resources within the OpenLAT collaboration, one of the youngest large-scale initiatives. Although this is a single case, it reflects a general pattern: the upfront cost of configuration generation supports a sustained physics programme through community reuse. In addition, a constant effort to develop novel algorithms and new software kernels is required to utilise state-of-the-art machines \cite{Finkenrath:2023sjg,Finkenrath:2024ptc}. An example for this ongoing effort is the development of \texttt{tmLQCD}, see fig.~\ref{fig:cost_alg_gpu}a, which could substantially speed up simulations using multi-grid solvers \cite{Alexandrou:2016izb}, as well as lately offloading solver and force computations to the \texttt{QUDA} library \cite{Kostrzewa:2022hsv,Alexandrou:2025bkm}. Another example is the recent porting of \texttt{HiRep} to GPU architectures \cite{DRACH2026110061}, which demonstrates excellent strong-scaling performance, see fig.~\ref{fig:cost_alg_gpu}b.
%\begin{figure}[t!]
%\centering
%\begin{subfigure}{0.46\textwidth}
%    \centering
%    \includegraphics[width=0.7\textwidth]{cost-3.pdf}
%    \caption{\texttt{tmLQCD}: Cost per molecular dynamics unit for a simulation with light, strange and charm quarks tuned to physical masses. Utilising multi-grid solvers sped up the simulations by a factor of 10. Using GPUs decreased the required node-hours by another factor of 10.}
%    \label{fig:cost_alg}
%\end{subfigure}
%\hfill
%\begin{subfigure}{0.46\textwidth}
%    \centering
%    \includegraphics[width=\textwidth]{hirep_scaling.pdf}
%    \caption{\texttt{HiRep}: Strong and weak scaling of the Dirac operator on LUMI-G with all dimensions parallelised up to 1024 GPUs for an SU(3) gauge group with fundamental fermions.}
%    \label{fig:cost_gpu}
%\end{subfigure}
%\end{figure}

\section{Main requirements for the future}
\subsection{Core hardware requirements for lattice investigations}
The performance of lattice applications is primarily determined by a combination of memory bandwidth, communication efficiency, and the ability to sustain performance across large parallel systems. The dominant computational kernel, namely the application of the Dirac operator within iterative solvers, is inherently bandwidth-bound. This situation has become even more pronounced with the adoption of novel algorithms, which reduce the overall floating-point workload while increasing pressure on memory access and data movement. \\
As a consequence:
\begin{itemize}[leftmargin=1.em,itemsep=0pt, parsep=0pt, topsep=0pt]
\item {\bfseries High memory bandwidth} is a critical requirement for future architectures. Equally important is the availability of low-latency, high-bandwidth interconnects, as lattice simulations rely heavily on nearest-neighbour communication patterns combined with frequent global reductions. Achieving efficient strong scaling to large fractions of modern supercomputers therefore depends both on network performance and the ability of the runtime system to overlap communication and computation effectively.
\item {\bfseries Large on-device memory}, such as high-bandwidth memory on accelerators, is also of central importance. Ma\-ny applications require storing multiple fields and propagators simultaneously, particularly during measurement phases where solutions of the Dirac equation are reused extensively. 
\item In this context, a {\bfseries balanced ratio between compute capability and memory bandwidth} is essential. Peak floating-point performance alone is not a reliable indicator of application performance; instead, sustained performance for stencil-like kernels with regular memory access patterns is the relevant metric for lattice workloads.
\item {\bfseries High and stable double precision performance} remains essential to ensure numerical accuracy, in particular for global reductions and long molecular-dynamics trajectories where rounding errors may accumulate. In selected cases, support for extended precision may also be required, for example in the presence of ill-conditioned systems or long summations. In the past this was usually realised in hardware, however, emulated high-precision arithmetic may be feasible, if performant and sufficiently verified.
\item Support for {\bfseries mixed-precision computations} is increasingly important. Modern solvers often exploit lower precision (such as single or half precision) in inner iterations to improve performance, while maintaining double precision for outer corrections and for the computation of physical observables.
% I split this point into 2.
\item Lattice QCD simulations require efficient exploitation of {\bfseries massive parallelism}. Architectures based on GPUs or many-core processors are well suited to this task, provided that lattice domains can be mapped efficiently onto the available parallel resources. Strong scaling to very large system sizes remains a key requirement, especially for large-volume simulations and long Markov chains. This, in turn, depends critically on both hardware capabilities and software support, including efficient MPI implementations and runtime systems. Given the regular structure of lattice discretisations, topology-aware network designs—such as {\bfseries hyper-torus or locality-aware networks} offer clear advantages by minimising communication overhead and preserving locality in nearest-neighbour exchanges.
\end{itemize}

\subsection{Software}
Given the high demand in computing resources, the lattice QCD community has a long and successful history in self-developing highly optimised software, often including hand-optimised code targeting specific HPC architectures.
For instance, the lattice community had already realised and applied the power of GPUs for scientific applications at a time when GPUs were mostly used for gaming \cite{Egri:2006zm}.
This has led to the development of a handful of simulation software suites, maintained by small groups of physicists often connected to one of the collaborations. Examples are \texttt{Chroma}, \texttt{Grid}, \texttt{HiRep}, \texttt{MILC-QCD}, \texttt{openQCD}, \texttt{QUDA}, \texttt{SIMULATeQCD}, and \texttt{tmLQCD}.

With the increasing complexity of modern computer architectures, performance portability has become a major challenge, as writing hand-optimised code for each platform is increasingly demanding. This has also sparked the development of novel approaches, for example implemented in \texttt{Kokkos}, a performance portability library that provides abstractions for parallel execution and data management. By expressing parallelism and memory access patterns at a high level, these approaches aim to enable compilers and backends to map the same code efficiently onto different architectures, such as multi-core CPUs and GPU accelerators, from a single source code base.

Maintaining such code suites requires continuous effort: continuous integration systems and automated performance regression tests are becoming increasingly important for sustaining reliability, reproducibility, and numerical efficiency. Moreover, these software packages require not only a stable and well maintained software stack, but also well-trained and sufficient personnel at the relevant computing centres: lattice QCD software often tests the boundaries of the hardware, and relies on the most recent features of the development toolchain. Therefore, small updates in the software stack regularly break the build chain of lattice QCD software, or lead to crashes. To address this, lattice QCD codes should be included into continuous testing pipelines of the high-performance computing centres. This is also useful for detecting hardware and compiler/module stack issues.
%added the last sentence

\subsection{Human resources}
Efficient utilisation of modern HPC systems requires close and sustained collaboration between research groups and computing centres. Performance analysis, architecture optimisation, and management of increasingly complex simulation workflows demand specialised expertise that goes beyond the traditional skills of individual research groups. 

Dedicated support personnel with expertise in parallel programming, GPU optimisation, and performance engineering are crucial for translating algorithmic requirements into efficient implementations on evolving HPC architectures.
%Dedicated support personnel with expertise in parallel programming, GPU optimisation, and performance engineering play a crucial role in enabling scientific applications to fully exploit modern HPC infrastructures. These experts act as an interface between domain scientists and hardware platforms, translating algorithmic requirements into efficient implementations and ensuring that applications remain performant as architectures evolve.
Such roles are particularly critical given that they require a combination of in-depth knowledge of numerical algorithms, hardware architectures, and software engineering practices, while at the same time being highly attractive for the broader job market. As a result, maintaining and developing this expertise within academia remains a structural challenge for many scientific communities, including lattice field theory, where software development is still largely distributed across individual groups and lacks a coherent, long-term European framework.

A successful model to address these challenges is provided by the Swiss Platform for Advanced Scientific Computing 
(PASC, \url{https://pasc-ch.org/about/index.html}). 
%(PASC)\footnote{\url{https://pasc-ch.org/about/index.html}}. 
PASC fosters close collaboration between universities and the Swiss National Supercomputing Centre, with a strong focus on application-driven software development and co-design between domain scientists and HPC experts. Extending similar frameworks at the European level would significantly strengthen the ability of lattice field theory and related disciplines to exploit current and future HPC systems effectively. In this context, the EuroHPC Centres of Excellence could represent a decisive step forward in establishing such coordinated efforts at the European scale.

\subsection{Critical view of present trends in EuroHPC}
European initiatives such as EuroHPC provide unprecedented computational capabilities and represent a major strategic investment for European science and technology. The deployment of large-scale systems across Europe has significantly expanded access to leadership-class computing resources. At the same time, several emerging trends offer new opportunities to further strengthen the impact of these infrastructures across a broad range of scientific domains.
%
%While machine-learning methods are increasingly useful within lattice QCD workflows, they do not replace the need for dedicated scientific HPC resources. %In particular, communication-bound, double-precision, strongly scaling workloads require architectural and procurement choices that are distinct from those optimised primarily for large-scale AI training.
%
\begin{itemize}[leftmargin=1.em,itemsep=0pt, parsep=0pt, topsep=0pt]
\item An important opportunity lies in broadening the criteria used to guide hardware procurement decisions. While current approaches capture peak floating-point performance and accelerator-driven workloads, complementing them with community-specific benchmarks would ensure a more balanced representation of scientific needs. This is not only relevant for lattice QCD, but also for other numerical approaches to problems in science and engineering, including hydrodynamics and numerical gravity \cite{PRACE2026ScientificInnovationCase}. These applications share demanding requirements in terms of communication costs, memory bandwidth, sustained strong scaling, network topology, latency, floating-point precision, and sustained bandwidth.
Incorporating representative workloads from these communities into benchmarking suites would therefore help ensure that future systems suit a broader range of scientific applications.
\item  A further strategic opportunity is the development of a clear long-term European roadmap towards post-exascale computing, covering both an intermediate five-year horizon and a longer ten-year perspective. Building on the strong foundation established by current EuroHPC systems, increased visibility on future architectures and infrastructure evolution would provide valuable guidance to scientific communities engaged in long-term code development and optimisation. Establishing such a roadmap, in alignment with other international efforts, would strengthen Europe’s position in the global HPC landscape and support sustained scientific leadership.
\item  The rapid growth of AI-driven workloads represents a major and welcome evolution in the HPC landscape, opening new scientific and technological opportunities. At the same time, this trend should be accompanied by a balanced allocation of resources that continues to support other computational science domains.
\item Established fields such as lattice field theory have historically been key drivers of innovation, producing advances in algorithms, software, and hardware whose broader impact is often difficult to anticipate but frequently transformative. Maintaining strong support for these research directions alongside emerging AI workloads will help preserve a diverse and resilient scientific ecosystem, maximising both immediate and long-term societal benefits.
\end{itemize}
A balanced approach to procurement, access policies, and long-term planning is essential for communication-bound applications such as lattice field theory.

\section{Conclusions}

Researchers in lattice field theory are among the largest and most efficient users of high-performance computing resources in fundamental science. The computational workloads place sustained demands on memory bandwidth, communication, and scalability, providing stringent benchmarks for current and future HPC architectures.
At the same time, lattice field theory illustrates the broader impact of curiosity-driven research. Its first-principles description of strongly interacting systems has driven methodological advances beyond its original domain, including the development of algorithms such as Hybrid Monte Carlo (Hamiltonian Monte Carlo) \cite{Duane:1987de} and contributions to the co-design of HPC systems, exemplified, e.g., by IBM's BlueGene architectures. These developments have had a lasting impact on computational science and data-driven methodologies.

The field remains one of the most computationally demanding areas of computational science, with progress tightly coupled to advances in algorithms, computing infrastructures and, potentially, machine learning. Looking beyond the exascale era, continued progress will require balanced HPC architectures capable of sustaining memory bandwidth and communication performance, together with robust and sustainable software ecosystems.

%A clear vision for post-exascale computing is therefore essential. This must ensure continued support for fundamental research domains that are not driven by current trends in AI and data-centric workloads, but which have historically been a major source of conceptual and technological innovation.
{\bf A clear vision for post-exascale computing is therefore essential}. This must ensure balanced support for fundamental research domains, such as lattice field theory, that drive conceptual advances and technological innovation, along with emerging trends. 

{\bf Sustained investment in human expertise is equally critical}. The increasing complexity of hardware and software demands dedicated support at the interface of physics, numerical methods, and HPC. Establishing stable career paths for research software engineers and strengthening collaboration between scientific communities and computing centres are essential components of an effective long-term strategy.

\section*{Acknowledgements}
The authors thank L. Del Debbio, A. Francis, B. Kostrzewa, C. Pica, S. Ryan, and S. Schaefer for valuable discussions and input. We also thank the OpenLAT collaboration for sharing information on their computational resource allocations. 
GA is supported by STFC grant ST/X000648/1 and a Royal Society Leverhulme Trust Senior Research Fellowship. JF~received funding from the European Research Council via the project
"LEEX" (GA 101170304).
CU and SK were supported by the Deutsche Forschungsgemeinschaft (DFG, German Research Foundation) as part of the CRC 1639 NuMeriQS – project no.\ 511713970 and under Germany’s Excellence Strategy – EXC 3107 – Project-ID 533766364 as part of the Cluster of Excellence Color-meets-Flavor.
GB, SK and CU are supported in part by DFG project 460248186 (PUNCH4NFDI).
We acknowledge EuroHPC JU for awarding access under the projects EHPC-EXT-2025E01-079 (222 knh on LUMI-G), EHPC-EXT-2023E01-010 (111 Mch on LUMI-C), EHPC-EXT-2025E01-091 (JUPITER), EHPC-EXT-2022E01-096 (120 Mch on LUMI-C), and EHPC-EXT-2023E01-018 (462 knh on Leonardo and 120 Mch on LUMI-C), which contributed to the studies presented.

\bibliography{bib}
\bibliographystyle{elsarticle-num}
\end{document}